\documentclass[final,5p,times,twocolumn]{elsarticle}

\usepackage{graphics}

\usepackage{amssymb}
\usepackage{amsthm}

\usepackage{textcomp}

\usepackage[nodots,nocompress]{numcompress}

\usepackage{lineno}

\biboptions{comma, square}

\journal{Nuclear Innstruments and Methods in Physics Research, A}

\begin{document}

\begin{frontmatter}



\title{On noise treatment in radio measurements of cosmic ray air showers}


\author[1]{F.G.~Schr\"oder\corref{cor1}}
\ead{frank.schroeder@kit.edu}
\author[1]{W.D.~Apel}
\author[2,14]{J.C.~Arteaga}
\author[3]{T.~Asch}
\author[4]{L.~B\"ahren}
\author[1]{K.~Bekk}
\author[5]{M.~Bertaina}
\author[6]{P.L.~Biermann}
\author[1,2]{J.~Bl\"umer}
\author[1]{H.~Bozdog}
\author[7]{I.M.~Brancus}
\author[8]{P.~Buchholz}
\author[4]{S.~Buitink}
\author[5,9]{E.~Cantoni}
\author[5]{A.~Chiavassa}
\author[1]{K.~Daumiller}
\author[2,15]{V.~de Souza}
\author[1]{P.~Doll} 
\author[1]{R.~Engel}
\author[4,10]{H.~Falcke} 
\author[1]{M.~Finger} 
\author[11]{D.~Fuhrmann}
\author[3]{H.~Gemmeke}
\author[8]{C.~Grupen}
\author[1]{A.~Haungs}
\author[1]{D.~Heck}
\author[4]{J.R.~H\"orandel}
\author[4]{A.~Horneffer}
\author[2]{D.~Huber}
\author[1]{T.~Huege}
\author[1,16]{P.G.~Isar}
\author[11]{K.-H.~Kampert}
\author[2]{D.~Kang}
\author[3]{O.~Kr\"omer}
\author[4]{J.~Kuijpers}
\author[4]{S.~Lafebre}
\author[2]{K.~Link}
\author[12]{P.~{\L}uczak}
\author[2]{M.~Ludwig}
\author[1]{H.J.~Mathes}
\author[2]{M.~Melissas}
\author[9]{C.~Morello}
\author[1]{S.~Nehls}
\author[1]{J.~Oehlschl\"ager}
\author[2]{N.~Palmieri}
\author[1]{T.~Pierog}
\author[11]{J.~Rautenberg}
\author[1]{H.~Rebel}
\author[1]{M.~Roth}
\author[3]{C.~R\"uhle}
\author[7]{A.~Saftoiu}
\author[1]{H.~Schieler}
\author[3]{A.~Schmidt}
\author[13]{O.~Sima}
\author[7]{G.~Toma}
\author[9]{G.C.~Trinchero}
\author[1]{A.~Weindl}
\author[1]{J.~Wochele}
\author[1]{M.~Wommer}
\author[12]{J.~Zabierowski}
\author[6]{J.A.~Zensus}

\cortext[cor1]{Corresponding author}

\address[1]{Karlsruhe Institute of Technology (KIT) - Campus North, Institut f\"ur Kernphysik, Germany}
\address[2]{Karlsruhe Institute of Technology (KIT) - Campus South, Institut f\"ur Experimentelle Kernphysik, Germany}
\address[3]{Karlsruhe Institute of Technology (KIT) - Campus North, Institut f\"ur Prozessdatenverarbeitung und Elektronik, Germany}
\address[4]{Radboud University Nijmegen, Department of Astrophysics, The Netherlands}
\address[5]{Dipartimento di Fisica Generale dell' Universita Torino, Italy}
\address[6]{Max-Planck-Institut f\"ur Radioastronomie Bonn, Germany}
\address[7]{National Institute of Physics and Nuclear Engineering, Bucharest, Romania}
\address[8]{Universit\"at Siegen, Fachbereich Physik, Germany}
\address[9]{INAF Torino, Istituto di Fisica dello Spazio Interplanetario, Italy}
\address[10]{ASTRON, Dwingeloo, The Netherlands}
\address[11]{Universit\"at Wuppertal, Fachbereich Physik, Germany}
\address[12]{Soltan Institute for Nuclear Studies, Lodz, Poland}
\address[13]{University of Bucharest, Department of Physics, Bucharest, Romania}

\address[14]{\scriptsize{now at: Universidad Michoacana, Instituto de F\'{\i}sica y Matem\'aticas, Mexico}}
\address[15]{\scriptsize{now at: Universidade S\~ao Paulo, Instituto de F\'{\i}sica de S\~ao Carlos, Brasil}}
\address[16]{\scriptsize{now at: Institute of Space Science, Bucharest, Romania}}

\begin{abstract}
Precise measurements of the radio emission by cosmic ray air showers require an adequate treatment of noise. Unlike to usual experiments in particle physics, where noise always adds to the signal, radio noise can in principle decrease or increase the signal if it interferes by chance destructively or constructively. Consequently, noise cannot simply be subtracted from the signal, and its influence on amplitude and time measurement of radio pulses must be studied with care. First, noise has to be determined consistently with the definition of the radio signal which typically is the maximum field strength of the radio pulse. Second, the average impact of noise on radio pulse measurements at individual antennas is studied for LOPES. It is shown that a correct treatment of noise is especially important at low signal-to-noise ratios: noise can be the dominant source of uncertainty for pulse height and time measurements, and it can systematically flatten the slope of lateral distributions. The presented method can also be transfered to other experiments in radio and acoustic detection of cosmic rays and neutrinos.
\end{abstract}


\end{frontmatter}


\section{Consistent definition of signal and noise}
Noise definitions applied so far in the field of cosmic ray radio detection are originating from communication engineering. There, a signal usually has a power much larger than the noise, and lasts for a time significantly longer than its oscillation period. Both is not true for air shower induced radio pulses. This has already been investigated in the frame of self-trigger development \cite{Kroemer08}, where the signal-to-noise ratio plays the role of a threshold. For data analysis, the situation is more complex because noise has to be defined consistently with the definition of the radio pulse height, which is the maximum of the field strength, in the case of LOPES \cite{HuegeArena2010}.

Independently of the specific signal and noise definitions, the following consistency criterion is demanded:

\begin{figure*}[t]
 \centering
 \includegraphics[width=0.47\linewidth]{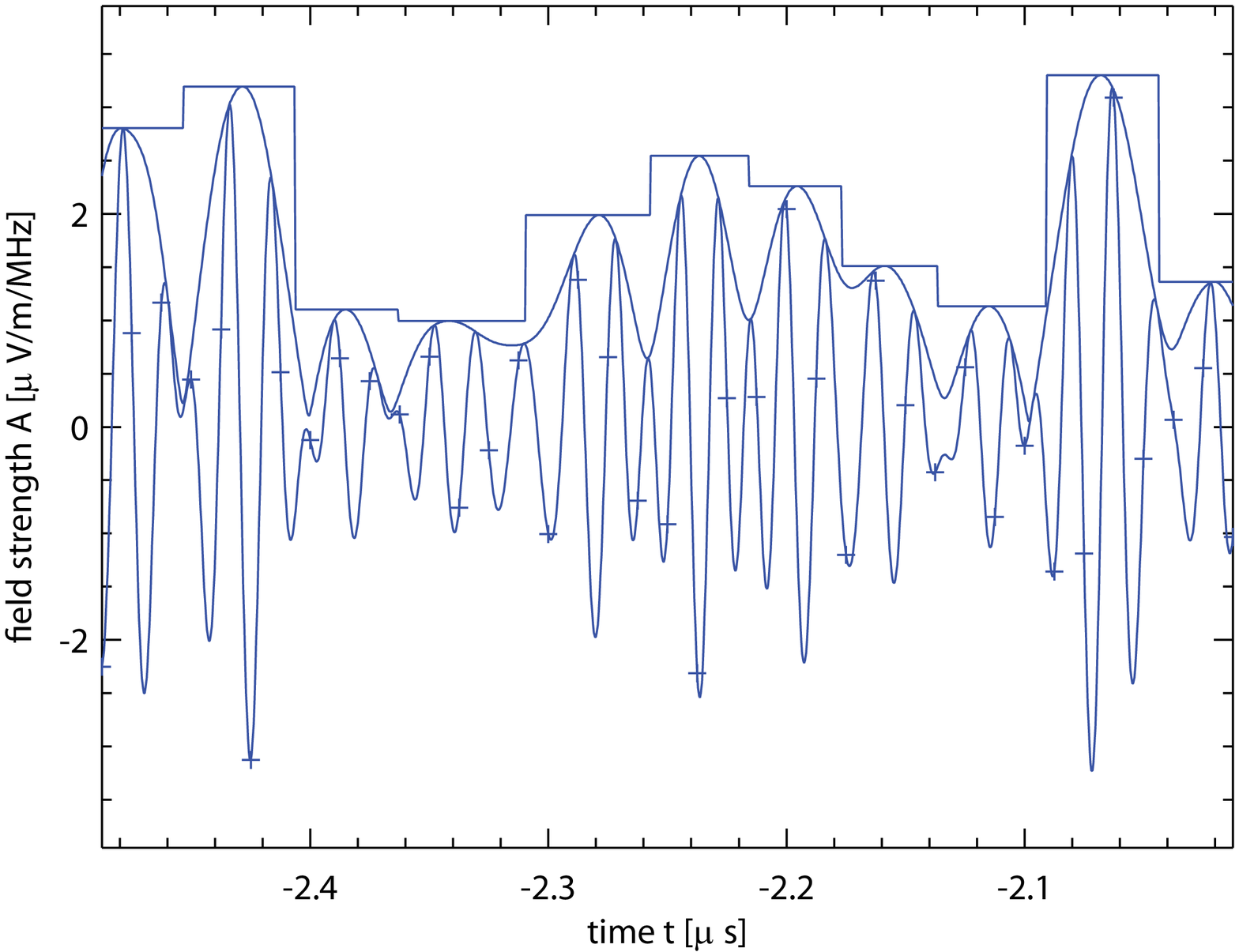}
 \hskip 0.04\linewidth
 \includegraphics[width=0.47\linewidth]{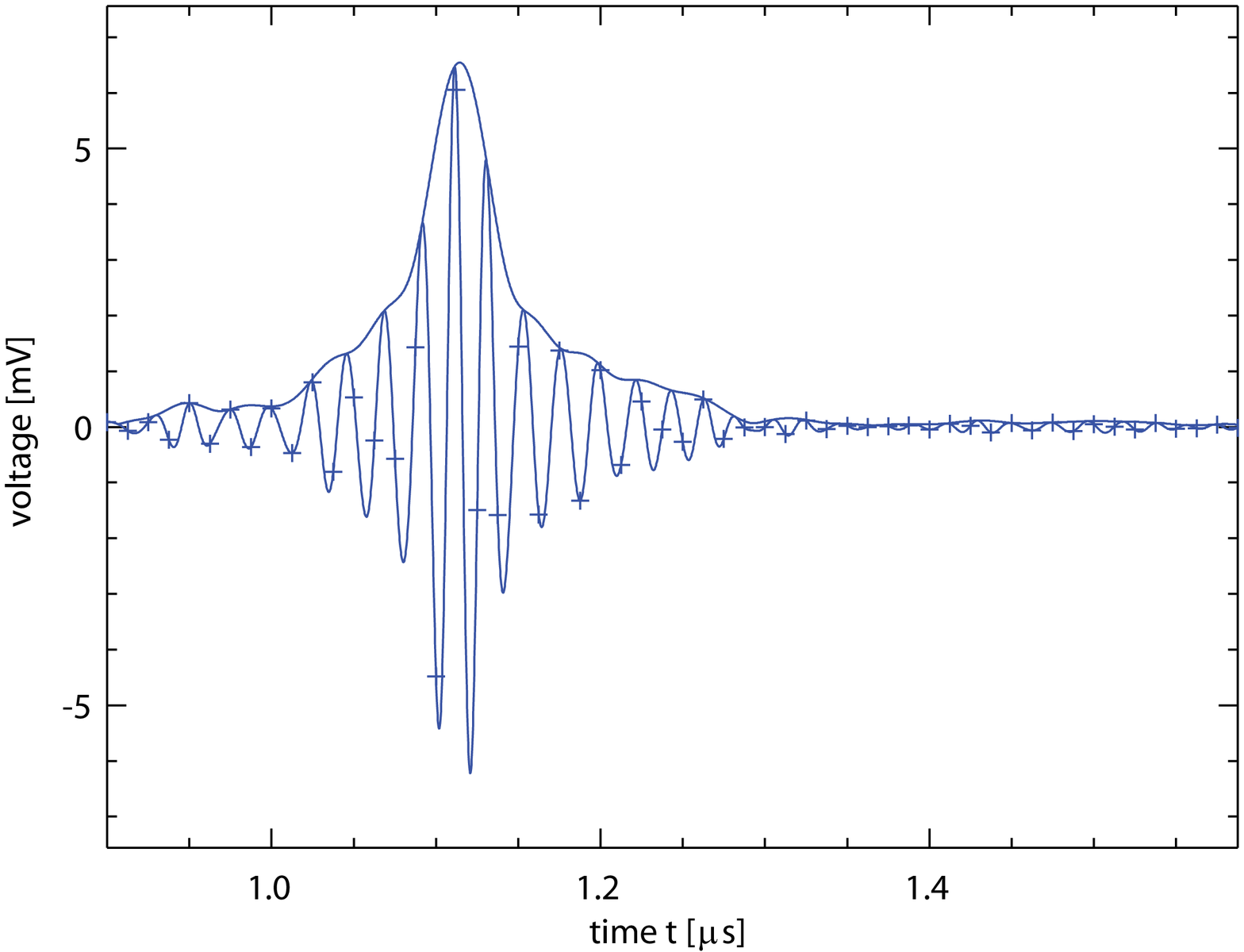}
 \caption{Typical noise measured with a LOPES antenna (left), and a test pulse from a pulse generator (right): sampled data points, the up-sampled trace and a Hilbert envelope of the trace are shown in both cases. The noise level is calculated as the weighted average of the local maxima of the envelope. This corresponds to the average level of the plotted step function with a step exactly in the middle between two local maxima.}
 \label{fig_noiseExample}
\end{figure*}

\begin{equation}
\textrm{for} ~~ {true~signal} = 0 ~~~ \longrightarrow ~~~ \frac{measured~signal}{noise} \stackrel{!}{=} 1
\end{equation}

The consistency criterion is supposed to hold only on average, because the noise level at the signal time can by chance be larger or smaller than the average noise level. In addition, even for a positive true signal, the measured signal-to-noise ratio can in some cases be smaller than one, since noise can interfere constructively or destructively with the air shower radio emission, and increase or decrease the measured signal compared to the true signal.

For LOPES, a consistent definition of signal and noise has been found for measurements at individual antennas, e.g., to reconstruct the lateral distribution \cite{Apel2010}. The signal is defined as the amplitude (field strength) of the radio pulse which is determined as the local maximum of an Hilbert envelope closest to the pulse time known from a preceding interferometric cross-correlation beam analysis (c.f. \cite{Horneffer07}). The noise level is defined as average amplitude in a time window ($10\,$\textmu s) before the radio pulse, and is calculated by the mean of all local maxima of the envelope. Because lower local maxima are more likely to have a smaller distance to neighboring maxima than higher maxima, it is necessary to weight each maximum with the distance to its neighbors when averaging (fig.~\ref{fig_noiseExample}, left). 

\begin{figure}
 \centering
 \includegraphics[width=0.9\linewidth]{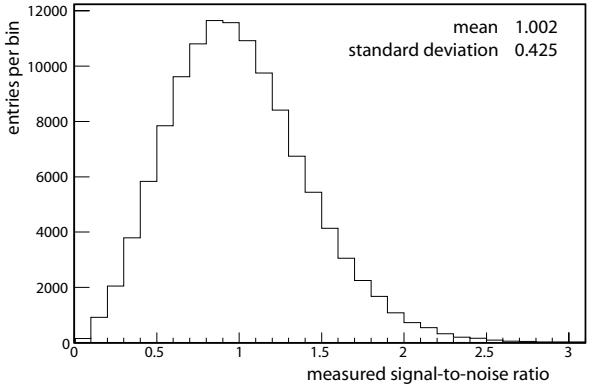}
 \caption{Signal-to-noise ratios of pure noise for a selection of LOPES events without signal.}
 \label{fig_SNRofNoise}
\end{figure}

It has been tested that these definitions of signal and noise do indeed fulfill the consistency criterion. With a selection of $200$ LOPES events without radio pulse, a large sample of $120,000$ noise intervals of $10\,$\textmu s width, each, has been obtained: the intervals are non-overlapping, cover different days and times of the day, as well as different antennas. The average signal-to-noise ratio of these intervals is compatible with $1$, as required (fig.~\ref{fig_SNRofNoise}).

With other definitions of noise, like the RMS of the field strength or its square (power), the mean of the absolute field strength or an unweighted mean of the local maxima of the envelope, the consistency criterion is not fulfilled. However, the ratio between the noise levels determined by different methods is constant within a few percent. Thus, results obtained with a different noise definition could be scaled to a consistent definition when accepting a small systematic error.

\section{Influence of noise on pulse height measurements}
The impact of noise on measurements of the pulse amplitude at individual antennas has been studied for LOPES with test pulses (fig.~\ref{fig_noiseExample}, right) of different width, and noise from real measurements. Therefore, the test pulses have been scaled with the LOPES analysis software to a certain amplitude $A_\textrm{true}$, and added to the noise intervals presented in the previous section. Afterwards, the measured signal height $A_\textrm{meas}$ can be obtained for each pulse, yielding a relation between the average true amplitude $A_\textrm{true}$ and the measured amplitude $A_\textrm{meas}$. To simplify the relation, all amplitudes have been normalized to the noise level, i.e., the noise level corresponds to $A=1$, and $A_\textrm{meas}$ is the measured signal-to-noise ratio.

\begin{figure}
 \centering
 \includegraphics[width=0.95\linewidth]{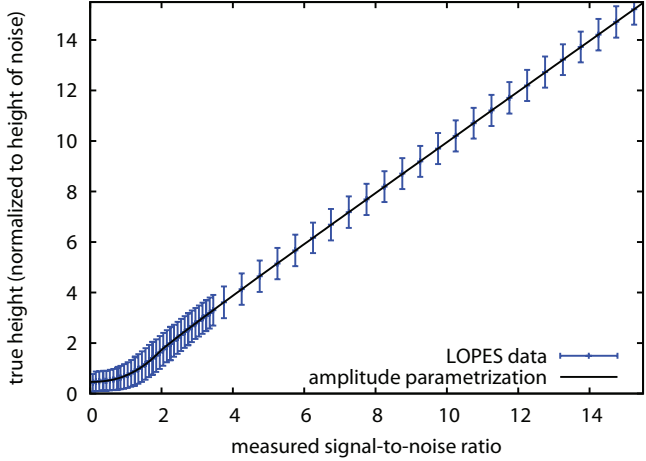}
 \caption{Relation between the true signal height $A_\textrm{true}$ and the measured signal-to-noise ratio $A_\textrm{meas}$. The error bars correspond to the standard deviations of the binned data, i.e. $\Delta A_\textrm{true}$, and not the the uncertainty of the plotted mean.}
 \label{fig_ampFormula}
\end{figure}

Because the real probability distribution of $A_\textrm{true}$ of air shower induced radio pulses is unknown, scaling factors for the test pulse heights have been generated for a flat distribution of $A_\textrm{true}$. As cross-check, also an exponentially decaying distribution has been tried, but the effect on the results is negligible.

To correct measured pulse amplitudes for the noise influence, the function $A_\textrm{true}(A_\textrm{meas})$ is required, which has been obtained by the following procedure. The test pulse data, which consist of $120,000$ samples with known $A_\textrm{true}$ and corresponding $A_\textrm{meas}$, have been sorted into bins. Thereby, each bin covers a certain interval of $A_\textrm{meas}$. The mean $A_\textrm{true}$ of each bin is then the average true amplitude corresponding to the measured amplitude $A_\textrm{meas}$ of the bin. At the same time, the standard deviation $\Delta A_\textrm{true}$ of each bin can be taken as error estimation of the true amplitude (fig.~\ref{fig_ampFormula}). Other methods to determine $A_\textrm{true}(A_\textrm{meas})$ failed. The inverse function of $A_\textrm{meas}(A_\textrm{true})$, which would be available directly, is not defined for $A_\textrm{meas} < 1$. Using confidence intervals instead of mean and standard deviation yields meaningless results for $A_\textrm{meas}$ close to $0$. More details about the method will be available in \cite{SchroederThesis}.

The validity of the relation $A_\textrm{true}(A_\textrm{meas})$ has been checked for various systematic effects, and no significant dependencies could be found. In detail, the following effects have been studied: the (up-)sampling rate, the shape of the test pulse, the antenna type and polarization used for noise measurement. Summarizing, any possible effects are negligible against the size of the error $\Delta A_\textrm{true}$, and against the calibration uncertainty due to environmental effects ($\sim 5\,\%$, c.f. \cite{Nehls08}).

The following parametrization has been found for $A_\textrm{true}$:
\begin{equation}
A_\textrm{true} = \sqrt{A_\textrm{meas}^2 - 1}  ~~~~~ \textrm{for} ~~ A_\textrm{meas} \gtrsim 2
\end{equation}
and for low signal-to-noise ratios:
\begin{equation}
A_\textrm{true} = a + b\cdot A_\textrm{meas}^c  ~~~~~ \textrm{for} ~~ A_\textrm{meas} \lesssim 2
\end{equation}
with $a = 0.4628 \pm 0.0066$, $b = 0.2491 \pm 0.0092$, and $c = 2.349 \pm 0.048$, determined by a fit, with a forced connection to the first formula at $A_\textrm{meas} = 2$.

This means that at high signal-to-noise ratios, the power of the measured signal is in average the sum of the power of the noise and the power of the radio pulse, i.e., noise generally is more likely to increase the signal amplitude than to decrease it. However, at low signal-to-noise ratios the behavior is not trivial. This demonstrates that a detailed study of the noise influence, like performed here for LOPES, is indeed necessary, especially because most LOPES events contain at least some antennas with signal-to-noise ratios $\lesssim 2$.

\begin{figure}
 \centering
 \includegraphics[width=0.95\linewidth]{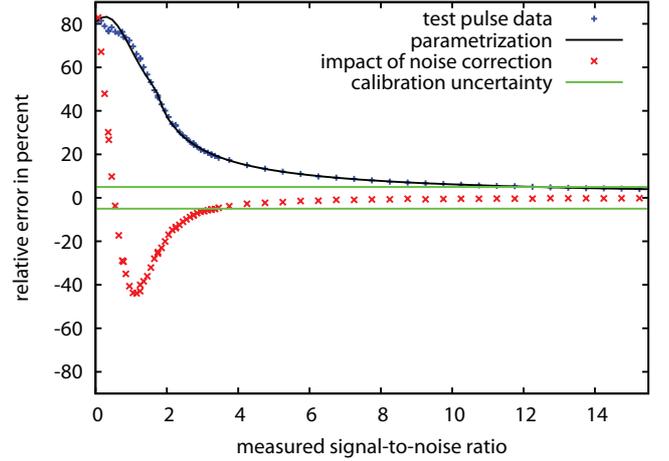}
 \caption{Relative uncertainty of the true pulse height ($\Delta A_\textrm{true}/A_\textrm{true}$, x), in each bin of the measured signal-to-noise ratio $A_\textrm{meas}$. This is compared with the error one would make, if the measured pulse amplitude would not be corrected for the noise influence ($(A_\textrm{true}-A_\textrm{meas})/A_\textrm{true}$, +), and with the calibration uncertainty due to environmental effects (horizontal lines).}
 \label{fig_relError}
\end{figure}

The noise dependent, statistical error of $A_\textrm{meas}$ which is determined as standard deviation $\Delta A_\textrm{true}$, like explained above, clearly exceeds the calibration uncertainty for signal-to-noise ratios $\lesssim 10$ (fig.~\ref{fig_relError}). Like for the amplitude itself, also the error $\Delta A_\textrm{meas}$ is parametrized differently for low and high signal-to-noise ratios:
\begin{equation}
\Delta A_\textrm{true} = d + e\cdot \exp{(+ A_\textrm{meas})} ~~~~ \textrm{for} ~~ A_\textrm{meas} \lesssim 1.68
\end{equation}
\begin{equation}
\Delta A_\textrm{true} = f + g\cdot \exp{(- A_\textrm{meas})} ~~~~ \textrm{for} ~~ A_\textrm{meas} \gtrsim 1.68
\end{equation}
with $d = 0.3103  \pm 0.0096$, $e = 0.0647 \pm 0.0029$, $f = 0.6162 \pm 0.0010$, and $g = 0.213 \pm 0.018$, obtained from a fit.

\begin{figure}
 \centering
 \includegraphics[width=0.97\linewidth]{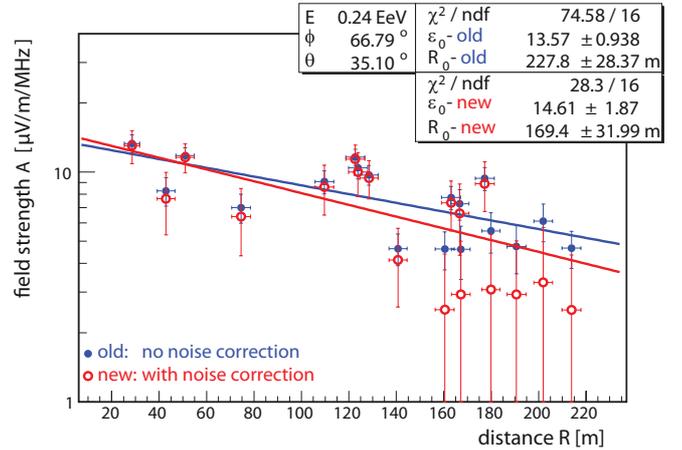}
 \caption{Example lateral distribution from a typical selection with and without correction of the noise influence.}
 \label{fig_exampleLatNoiseInfluence}
\end{figure}

\section{Influence on lateral distributions}
Since the amount by which noise in average increases or decreases the amplitude of radio pulses depends on the signal-to-noise ratio, it also has an impact on the slope of lateral distributions of the air shower induced radio emission. For a typical selection of LOPES events (e.g., the one used in \cite{Apel2010}), noise significantly flattens lateral distributions (fig.~\ref{fig_exampleLatNoiseInfluence}), except for events with very high signal-to-noise ratios at all antennas.

If an exponential function $A(R) = \epsilon_0 \cdot \exp(R/R_0)$ is fitted to each lateral distribution with a slope parameter $R_0$, noise typically increases $R_0$ by $10-20\,\%$, which is not negligible compared to other uncertainties. The effect on the amplitude parameter is smaller and vanishes only for events with high signal-to-noise ratio at all antennas. Fortunately, it is now possible to correct for the influence of noise in every single measurement, at each individual antenna, by the parametrization formulas presented above, to obtain the 'true' lateral distributions.

\begin{figure}
 \centering
 \includegraphics[width=0.95\linewidth]{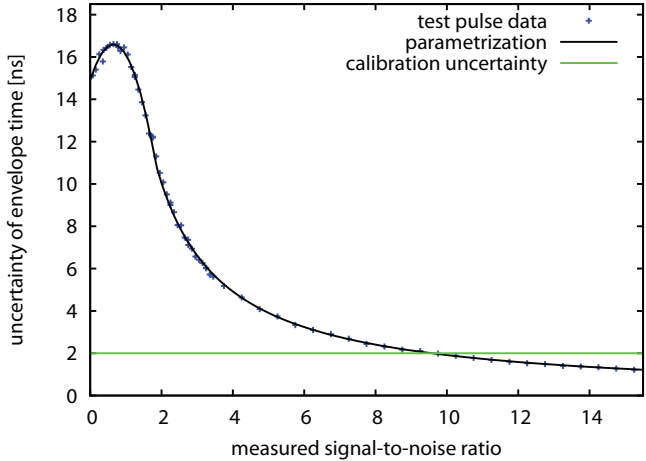}
 \caption{Average deviation between the measured time of the envelope maximum after adding noise, and the original time before adding noise,  and the time calibration uncertainty as reference.}
 \label{fig_timeFormula}
\end{figure}

\section{Influence on pulse arrival time measurements}
The influence of noise on the measurement of pulse arrival times has been studied in a similar way like the influence on pulse amplitudes, defining the pulse arrival time at an antenna as the time when the pulse amplitude is measured. Noise randomly shifts the true pulse arrival time of test pulses to earlier or later times, and no significant tendency to either one is observed.

The mean of the absolute shifts of the pulse time at a certain signal-to-noise ratio is defined as the noise dependent time uncertainty $\Delta t$ (fig.~\ref{fig_timeFormula}). For signal-to-noise ratios $\lesssim 10$, it can be much larger than the time calibration uncertainty, which is about $2\,$ns for pulse arrival time measurements at individual antennas \cite{Schroeder2010}. For the interferometric cross-correlation beam analysis used in LOPES, the impact of noise will be smaller, because shifts at different antennas will average out, but a quantitative study is beyond the scope of this paper.

Several possible systematic effects have been examined. As expected, $\Delta t$ does neither depend on the antenna type nor its polarization. It does depend on the sampling frequency, but the effect becomes negligible against the time calibration uncertainty of LOPES, if data are up-sampled to at least $640\,$MHz. Unfortunately, $\Delta t$ depends on the pulse shape, but no correlation with easy accessible parameters like pulse height or width could be found. For this reason, and since it is unknown which test pulse shape does best describe the real cosmic ray induced pulses, the following parametrization of $\Delta t$ takes into account the average behavior of all tested pulse shapes:
\begin{equation}
\Delta t = A_\textrm{meas}^{-1.03} \cdot 20.5\,\textrm{ns} ~~~~ \textrm{for} ~~ A_\textrm{meas} \gtrsim 1.8
\end{equation}
whereby $A_\textrm{meas}$ is the measured signal-to-noise ratio, and all parameters have been adjusted by hand to fit the data. The arrival time uncertainty $\Delta t$ does not become arbitrary large for low signal amplitudes, because the time interval for signal search depends on the preceding cross-correlation beam-forming. For high signal-to-noise ratios, the behavior is consistent with experience from beam-forming where the resolution improves with the signal-to-noise ratio.

\section{Conclusions}
Treating noise correctly in measurements of radio pulses emitted by air showers is especially tricky at low signal amplitudes. Nevertheless, it is important because events close to the detection threshold will always contain antennas with low signal-to-noise ratios. For instance, it has been shown that noise systematically flattens lateral distributions measured with LOPES.

Alternatively, only events at high signal-to-noise ratios could be studied, where the noise influence becomes negligible against calibration uncertainties. Then, the signal could be defined as the integrated power of the radio pulse, if the integration time is large against the pulse width. This is a tempting approach, because this method could be realized directly in analog electronics, and would allow cheaper and easier designs of radio detectors. However, this will be paid with a higher detection threshold, and is no option for current experiments like LOPES or AERA \cite{Berg09}.

Independently of the experimental design, attention must be paid to define signal and noise consistently in any analyses. Consistent definitions have been presented for LOPES for pulse measurements at individual antennas. A corresponding study of the noise influence on beam-forming analyses is more complex, and might better be performed with simulated pulses. Nevertheless, the presented method and results for the noise impact at individual antennas can probably be transfered to any experiment based on radio or acoustic arrays, where the signal consists of a short, bandwidth limited pulse, and noise can interfere in both ways, destructively and constructively with the signal.


\end{document}